

\font\bigbold=cmbx12
\font\ninerm=cmr9
\font\eightrm=cmr8
\font\sixrm=cmr6
\font\fiverm=cmr5
\font\ninebf=cmbx9
\font\eightbf=cmbx8
\font\sixbf=cmbx6
\font\fivebf=cmbx5
\font\ninei=cmmi9  \skewchar\ninei='177
\font\eighti=cmmi8  \skewchar\eighti='177
\font\sixi=cmmi6    \skewchar\sixi='177
\font\fivei=cmmi5
\font\ninesy=cmsy9 \skewchar\ninesy='60
\font\eightsy=cmsy8 \skewchar\eightsy='60
\font\sixsy=cmsy6   \skewchar\sixsy='60
\font\fivesy=cmsy5
\font\nineit=cmti9
\font\eightit=cmti8
\font\ninesl=cmsl9
\font\eightsl=cmsl8
\font\ninett=cmtt9
\font\eighttt=cmtt8
\font\tenfrak=eufm10
\font\ninefrak=eufm9
\font\eightfrak=eufm8
\font\sevenfrak=eufm7
\font\fivefrak=eufm5
\font\tenbb=msbm10
\font\ninebb=msbm9
\font\eightbb=msbm8
\font\sevenbb=msbm7
\font\fivebb=msbm5
\font\tensmc=cmcsc10


\newfam\bbfam
\textfont\bbfam=\tenbb
\scriptfont\bbfam=\sevenbb
\scriptscriptfont\bbfam=\fivebb
\def\Bbb{\fam\bbfam}

\newfam\frakfam
\textfont\frakfam=\tenfrak
\scriptfont\frakfam=\sevenfrak
\scriptscriptfont\frakfam=\fivefrak
\def\frak{\fam\frakfam}

\def\smc{\tensmc}


\def\eightpoint{%
\textfont0=\eightrm   \scriptfont0=\sixrm
\scriptscriptfont0=\fiverm  \def\rm{\fam0\eightrm}%
\textfont1=\eighti   \scriptfont1=\sixi
\scriptscriptfont1=\fivei  \def\oldstyle{\fam1\eighti}%
\textfont2=\eightsy   \scriptfont2=\sixsy
\scriptscriptfont2=\fivesy
\textfont\itfam=\eightit  \def\it{\fam\itfam\eightit}%
\textfont\slfam=\eightsl  \def\sl{\fam\slfam\eightsl}%
\textfont\ttfam=\eighttt  \def\tt{\fam\ttfam\eighttt}%
\textfont\frakfam=\eightfrak \def\frak{\fam\frakfam\eightfrak}%
\textfont\bbfam=\eightbb  \def\Bbb{\fam\bbfam\eightbb}%
\textfont\bffam=\eightbf   \scriptfont\bffam=\sixbf
\scriptscriptfont\bffam=\fivebf  \def\bf{\fam\bffam\eightbf}%
\abovedisplayskip=9pt plus 2pt minus 6pt
\belowdisplayskip=\abovedisplayskip
\abovedisplayshortskip=0pt plus 2pt
\belowdisplayshortskip=5pt plus2pt minus 3pt
\smallskipamount=2pt plus 1pt minus 1pt
\medskipamount=4pt plus 2pt minus 2pt
\bigskipamount=9pt plus4pt minus 4pt
\setbox\strutbox=\hbox{\vrule height 7pt depth 2pt width 0pt}%
\normalbaselineskip=9pt \normalbaselines
\rm}


\def\ninepoint{%
\textfont0=\ninerm   \scriptfont0=\sixrm
\scriptscriptfont0=\fiverm  \def\rm{\fam0\ninerm}%
\textfont1=\ninei   \scriptfont1=\sixi
\scriptscriptfont1=\fivei  \def\oldstyle{\fam1\ninei}%
\textfont2=\ninesy   \scriptfont2=\sixsy
\scriptscriptfont2=\fivesy
\textfont\itfam=\nineit  \def\it{\fam\itfam\nineit}%
\textfont\slfam=\ninesl  \def\sl{\fam\slfam\ninesl}%
\textfont\ttfam=\ninett  \def\tt{\fam\ttfam\ninett}%
\textfont\frakfam=\ninefrak \def\frak{\fam\frakfam\ninefrak}%
\textfont\bbfam=\ninebb  \def\Bbb{\fam\bbfam\ninebb}%
\textfont\bffam=\ninebf   \scriptfont\bffam=\sixbf
\scriptscriptfont\bffam=\fivebf  \def\bf{\fam\bffam\ninebf}%
\abovedisplayskip=10pt plus 2pt minus 6pt
\belowdisplayskip=\abovedisplayskip
\abovedisplayshortskip=0pt plus 2pt
\belowdisplayshortskip=5pt plus2pt minus 3pt
\smallskipamount=2pt plus 1pt minus 1pt
\medskipamount=4pt plus 2pt minus 2pt
\bigskipamount=10pt plus4pt minus 4pt
\setbox\strutbox=\hbox{\vrule height 7pt depth 2pt width 0pt}%
\normalbaselineskip=10pt \normalbaselines
\rm}


\def\pagewidth#1{\hsize= #1}
\def\pageheight#1{\vsize= #1}
\def\hcorrection#1{\advance\hoffset by #1}
\def\vcorrection#1{\advance\voffset by #1}

\newif\iftitlepage   \titlepagetrue               
\newtoks\titlepagefoot     \titlepagefoot={\hfil} 
\newtoks\otherpagesfoot    \otherpagesfoot={\hfil\tenrm\folio\hfil}
\footline={\iftitlepage\the\titlepagefoot\global\titlepagefalse
           \else\the\otherpagesfoot\fi}

\font\extra=cmss10 scaled \magstep0
\setbox1 = \hbox{{{\extra R}}}
\setbox2 = \hbox{{{\extra I}}}
\setbox3 = \hbox{{{\extra C}}}
\setbox4 = \hbox{{{\extra Z}}}
\setbox5 = \hbox{{{\extra N}}}

\def\RRR{{{\extra R}}\hskip-\wd1\hskip2.0 
   true pt{{\extra I}}\hskip-\wd2
\hskip-2.0 true pt\hskip\wd1}
\def\Real{\hbox{{\extra\RRR}}}    

\def\CCC{{{\extra C}}\hskip-\wd3\hskip 2.5 true pt{{\extra I}}
\hskip-\wd2\hskip-2.5 true pt\hskip\wd3}
\def\Complex{\hbox{{\extra\CCC}}}   

\def\ZZZ{{{\extra Z}}\hskip-\wd4\hskip 2.5 true pt{{\extra Z}}}
\def\Zed{\hbox{{\extra\ZZZ}}}       




\def\Z{{\Zed}}
\def\R{{\Real}}
\def\C{{\Complex}}

\def\pa{\partial}

\def\tr{{\rm Tr}\,}
\def\pa{\partial}
\def\nvec{{\bf n}}
\def\conf{{\cal Q}}

\def\cpone{\C{\!\!}P^1}

\def\qg{\widehat{g}}
\def\cL{\widehat{{\cal L}}}
\def\SO{\widehat{U}(g_{21})}


\def\abstract#1{{\parindent=30pt\narrower\noindent\ninepoint\openup
2pt #1\par}}


\newcount\notenumber\notenumber=1
\def\note#1
{\unskip\footnote{$^{\the\notenumber}$}
{\eightpoint\openup 1pt #1}
\global\advance\notenumber by 1}


\def\frac#1#2{{#1\over#2}}

\def\({\left(}
\def\){\right)}
\def\<{\langle}
\def\>{\rangle}

\def\pmb#1{\setbox0=\hbox{$#1$}%
   \kern-.025em\copy0\kern-\wd0
   \kern.05em\copy0\kern-\wd0
   \kern-.025em\raise.0433em\box0 }


\global\newcount\secno \global\secno=0
\global\newcount\meqno \global\meqno=1
\global\newcount\appno \global\appno=0
\newwrite\eqmac
\def\romappno{\ifcase\appno\or A\or B\or C\or D\or E\or F\or G\or H
\or I\or J\or K\or L\or M\or N\or O\or P\or Q\or R\or S\or T\or U\or
V\or W\or X\or Y\or Z\fi}
\def\eqn#1{
        \ifnum\secno>0
       \eqno(\the\secno.\the\meqno)\xdef#1{\the\secno.\the\meqno}
        \else\ifnum\appno>0
  \eqno({\rm\romappno}.\the\meqno)\xdef#1{{\rm\romappno}.\the\meqno}
          \else
            \eqno(\the\meqno)\xdef#1{\the\meqno}
          \fi
        \fi
\global\advance\meqno by1 }


\global\newcount\refno
\global\refno=1 \newwrite\reffile
\newwrite\refmac
\newlinechar=`\^^J
\def\ref#1#2{\the\refno\nref#1{#2}}
\def\nref#1#2{\xdef#1{\the\refno}
\ifnum\refno=1\immediate\openout\reffile=refs.tmp\fi
\immediate\write\reffile{
     \noexpand\item{[\noexpand#1]\ }#2\noexpand\nobreak}
     \immediate\write\refmac{\def\noexpand#1{\the\refno}}
   \global\advance\refno by1}
\def\semi{;\hfil\noexpand\break ^^J}
\def\nl{\hfil\noexpand\break ^^J}
\def\refn#1#2{\nref#1{#2}}

\def\ijmp#1#2#3{{\sl Int.  J.  Mod.  Phys.} {\bf A{#1}} (19{#2}) #3}

\def\np#1#2#3{{\sl Nucl.  Phys.} {\bf B{#1}} (19{#2}) #3} 

\def\plB#1#2#3{{\sl Phys.  Lett.} {\bf {#1}B} (19{#2}) #3} 
 
\def\prB#1#2#3{{\sl Phys.  Rev.} {\bf B{#1}} (19{#2}) #3} 
\def\prD#1#2#3{{\sl Phys.  Rev.} {\bf D{#1}} (19{#2}) #3} 
\def\prl#1#2#3{{\sl Phys.  Rev.  Lett.} {\bf #1} (19{#2}) #3} 

\def\ptp#1#2#3{{\sl Prog. Theor. Phys.} {\bf {#1}} (19{#2}) #3}
\def\rmp#1#2#3{{\sl Rev.  Mod.  Phys.} {\bf {#1}} (19{#2}) #3} 
\def\zpc#1#2#3{{\sl Z.  Phys.} {\bf {#1}} (19{#2}) #3} 


\refn\BKW
{M.J.\ Bowick, D.\ Karabali, L.C.R.\ Wijewardhana,
\np{271}{86}{417}.
}

\refn\SS
{G.W.\ Semenoff, P.\ Sodano,
\np{328}{89}{753}.
}

\refn\F
{
S.\ Forte,
\rmp{64}{92}{193}.
}

\refn\SKKR
{
S.L.\ Sondhi, A.\ Karlhede, S.A.\ Kivelson, E.H.\ Rezayi,
\prB{47}{93}{16419}.
}

\refn\AB
{
W.\ Apel, Y.A.\ Bychkov,
\prl{78}{97}{2188}; \prl{79}{97}{3792}.
}

\refn\VY
{
G.E.\ Volovik, V.M.\ Yakovenko,
\prl{79}{97}{3791}; cond-mat/9711076.
}


\refn\WZ
{F.\ Wilczek, A.\ Zee,
\prl{51}{83}{2250}.
}

\refn\PRS
{
P.K.\ Panigrahi, S.\ Roy, W.\ Scherer,
\prl{61}{88}{2827}; \prD{38}{88}{3199}.
}

\refn\BSD
{
M.\ Bergeron, G.W.\ Semenoff, R.R.\ Douglas,
\ijmp{7}{92}{2417}.
}

\refn\OS
{
H.\ Otsu, H.\ Sato,
\ptp{91}{94}{1199}; \zpc{C64}{94}{177}.
}

\refn\CM
{
B.\ Chakraborty, A.S.\ Majumdar,
{hep-th/9710028}.
}

\refn\LRV
{
T.\ Lee, C.N.\ Rao, K.S.\ Viswanathan,
\prD{39}{89}{2350}.
}

\refn\HST
{
Z.\ Hlousek, D.\ S\'en\'echal, S.-H.H.\ Tye,
\prD{41}{90}{3773}.
}

\refn\PP
{
N.K.\ Pak, R.\ Percacci,
\prD{43}{91}{1375}.
}

\refn\KTT
{
H.\ Kobayashi, S.\ Tanimura, I.\ Tsutsui, 
KEK Preprint 97-19, hep-th/9705183, to appear in 
{\sl Nucl.  Phys.} {\bf B}.
}

\refn\BST
{
A.P.\ Balachandran, A.\ Stern, G.\ Trahern,
\prD{19}{79}{2416}.
}

\refn\BMSS
{
A.P.\ Balachandran, G.\ Marmo, B.S.\ Skagerstam, A.\ Stern,
\lq\lq Classical Topology and Quantum States\rq\rq,
World Scientific, Singapore, 1991.
}

\refn\WZa
{
Y.S.\ Wu, A.\ Zee,
\plB{147}{84}{325};
}

\refn\K
{
D.\ Karabali,
\ijmp{6}{91}{1369}.
}

\refn\WZb
{
Y.S.\ Wu, A.\ Zee,
\np{272}{86}{322}.
}

\refn\DH
{
E.\ D'Hoker,
\plB{357}{95}{539}.
}

\refn\JW
{
R.\ Jackiw, E.J.\ Weinberg,
\prl{64}{90}{2234}.
}


\pageheight{23cm}
\pagewidth{14.8cm}
\hcorrection{0mm}
\magnification= \magstep1
\def\bsk{%
\baselineskip= 16.8pt plus 1pt minus 1pt}
\parskip=5pt plus 1pt minus 1pt
\tolerance 6000



\null

{
\leftskip=100mm
\hfill\break
KEK Preprint 97-241
\hfill\break
ICRR-Report-404-97-27
\hfill\break
\par}

\smallskip
\vfill
{\baselineskip=18pt

\centerline{\bigbold 
Hopf Term, Fractional Spin and Soliton Operators}
\centerline{\bigbold
in the O(3) Nonlinear Sigma Model}
 
\vskip 30pt

\centerline{
\smc Masaomi Kimura
\note
{E-mail:\quad masaomi@icrr.u-tokyo.ac.jp}
}

\vskip 5pt
{
\baselineskip=13pt
\centerline{\it Institute for Cosmic Ray Research}
\centerline{\it University of Tokyo}
\centerline{\it Midori, Tanashi, Tokyo 188, Japan}
}

\vskip 15pt

\centerline{
\smc 
Hiroyuki Kobayashi\note
{E-mail:\quad kobayasi@tanashi.kek.jp}
\quad {\rm and} \quad 
Izumi Tsutsui\note
{E-mail:\quad tsutsui@tanashi.kek.jp}
}

\vskip 5pt

{
\baselineskip=13pt
\centerline{\it Institute of Particle and Nuclear Studies}
\centerline{\it High Energy Accelerator Research 
            Organization (KEK), Tanashi Branch}
\centerline{\it Midori, Tanashi, Tokyo 188, Japan}
}

\vskip 50pt

\abstract{%
{\bf Abstract.}\quad
We re-examine three issues, the Hopf term, fractional
spin and the soliton operators, 
in the 
$2 + 1$ dimensional $O(3)$ nonlinear sigma model based
on the adjoint orbit parametrization (AOP) introduced earlier.
It is shown that the Hopf term is well-defined for 
configurations of any soliton charge $Q$ if we adopt
a time independent boundary condition at spatial infinity.
We then develop the Hamiltonian formulation of the model
in the AOP and thereby argue that 
the well-known $Q^2$-formula for fractional 
spin holds only for a restricted
class of configurations.  Operators which create states of
given classical configurations of any soliton number
in the (physical) Hilbert space are constructed. 
Our results clarify some of the points which are crucial for
the above three topological issues and yet have 
remained obscure in the literature.
}

\vfill\eject

\bsk


\secno=1 \meqno=1 

\noindent
{\bf 1. Introduction}
\medskip

The $O(3)$ nonlinear sigma model (NSM) describes 
physical systems that 
undergo a spontaneous breakdown of the global 
symmetry $O(3)$.  It is probably one of the most 
widely-applicable field
theory models, being used in fields ranging from
condensed matter physics to high energy physics.
The model was studied
intensively around a decade ago when the theoretical
possibility of particles possessing
a fractional spin and statistics in $2 + 1$
dimensions, namely anyons, 
attracted a lot of attention [\BKW, \SS] 
(see also [\F] for a review) in expectation of 
possible relevance to
the (fractional) quantum Hall effect [\SKKR, \AB, \VY] 
and high-$T_{\rm c}$ superconductivity.
Such a phenomenon in the NSM was originally 
suggested by Wilczek and Zee [\WZ] under the presence
of a topological term, the Hopf term, which endows solitons 
(Skyrmions) admitted by the model with a nontrivial phase 
factor under space rotation or interchange.
Since then, 
several authors have examined 
the NSM in order, for example, 
to furnish a firmer 
basis for fractional 
spin and statistics [\PRS, \BSD, \OS, \CM], and to 
explore its
possible extensions/modifications [\LRV, \HST, \PP] 
allowing for anyons.
Recently, we added  
to this series of investigations a study [\KTT] 
of the NSM using the adjoint orbit parametrization (AOP),
which has been known [\BST, \BMSS, \PP] for some time but
not thoroughly utilized so far,
unlike the other familiar $\cpone$ parametrization [\PRS].
The aim of this paper
is to present a complete AOP analysis of the NSM and
thereby re-examine the three topological issues, the Hopf term,
fractional spin and soliton operators, discussed previously.
We shall find that some of the points which are important 
for the topological phenomenon to occur and yet have 
remained obscure are clarified in the AOP.

To understand the points we are going to address in this paper, 
let us recall that the $O(3)$ nonlinear sigma model is a system of 
spin vectors ${\bf n}(x) = (n_1(x), n_2(x), n_3(x))$ 
constrained on the 2-sphere,
${\bf n}^2(x) = \sum_a n_a^2(x) = 1$.
In three dimensions the system is governed by the action, 
$$
I_0 =  \int d^3 x \, {1 \over{2\lambda^2}}\, 
\pa_\mu {\bf n}(x)\cdot
\pa^\mu {\bf n}(x),
\eqn\naction
$$
where $\lambda$ is a coupling constant.  We take the 
spacetime to be $\R^2 \times [0, T]$ and, as usual, assume
that the spin vectors approach
a constant vector at spatial infinity,
$$
{\bf n}(x) = {\bf n}({\bf x}, t)
\longrightarrow {\bf n}(\infty)
\qquad \hbox{as} \quad \Vert {\bf x} \Vert \rightarrow \infty,
\eqn\bdcon
$$
at all times $t \in [0, T]$.  
This boundary condition allows us to 
regard $\nvec(x)$ as a map $S^2 \times [0, T] \rightarrow S^2$
by identifying all points
at spatial infinity of $\R^2$ 
to be the south-pole of the sphere $S^2$ which is 
compactified from $\R^2$. 
The configuration space of the model is then given 
by the space of these maps (at a fixed time),
$$
\conf = {\rm Map}_0(S^2, S^2),
\eqn\cfsp
$$
where the subscript 0 indicates 
that the space consists of based maps ({\it i.e.}, those 
with the image of the south-pole fixed).

The topological structure of the model may be
characterized by the homotopy groups of the configuration
space.  Using the identity $\pi_n(\conf) = \pi_{n+2}(S^2)$
which holds for 
the space of based maps $\conf$ (see, {\it e.g.}, [\KTT]), 
we find
$$
\pi_0(\conf) = \pi_{2}(S^2) = \Z.
\eqn\pizero
$$
This implies that the space $\conf$
splits into disconnected sectors characterized by
an integer.  This integer is the soliton number
$Q := \int_{S^2}d^2x\, J^0(x)$ which is the charge of the
conserved topological current   
$$
J^\mu = {1\over{8\pi}}
\epsilon^{\mu\nu\lambda}\epsilon_{abc}\, n_a
\pa_\nu n_b\pa_\lambda n_c.
\eqn\topcurrent
$$
We also find
$$
\pi_1(\conf) = \pi_{3}(S^2) = \Z,
\eqn\pione
$$
which shows that, in each sector, configurations are 
characterized by another integer, called the Hopf
(or instanton) number.  To express 
the Hopf number as a term in the action, conventionally
one considers [\WZ]
$$
H = - \int d^3 x\, A_\mu(x)\, J^\mu(x),
\eqn\quasihopf
$$
with the vector potential $A_\mu$ being 
defined from the current
by the relation, 
$J^\mu = \epsilon^{\mu\nu\lambda}\pa_\nu A_\lambda$.
The expression (\quasihopf)
is expected to serve as a topological term 
expressing the Hopf number and 
--- if it is well-defined to any spin vectors --- 
may be added to the action
as $I = I_0 + \theta \, H$ with $\theta$ an angle parameter.
(The topological term can also be induced either by interactions with
fermions [\HST] or by a pure quantum mechanical 
topological effect [\WZa, \KTT], even though it does not appear
at the classical level.)
The problem, however, is that the expression
(\quasihopf) can reproduce the Hopf number
only for configurations in the zero soliton sector.  
This is obvious because, if the original 
map $S^2 \times [0, T] \rightarrow S^2$ is to be 
regarded as a map $S^3 \rightarrow S^2$ 
by means of suspension $S\cdot S^2 \simeq S^3$ ({\it i.e.}, by
contracting the sphere $S^2$ at $t = 0$
and $t = T$), it must be homotopic to a constant map at the
both ends of the period $[0, T]$ and hence
must have a vanishing soliton number.

In our previous paper [\KTT] we employed 
the AOP for the spin vectors, and argued
a possible definition of the Hopf term to solitons,\note{
In this paper we use the term
`soliton' to indicate a configuration which has a nonvanishing 
soliton number,
not just a soliton (topologically nontrivial) 
solution of the equations of motion.}
that is, the topological term which reproduces the Hopf number
to configurations of any soliton numbers.  
The idea is simply to convert the
configuration of non-vanishing 
soliton number to a corresponding one of vanishing 
soliton number using a fixed (standard)
configuration which has the opposite soliton number.
In this paper we first show in sect.2 that, in the AOP,
there exists a boundary condition for the field  
such that the above conversion
procedure becomes unnecessary.  We shall see that,
although the boundary condition appears slightly more 
restrictive 
than the one naively expected from (\bdcon), it is  
actually the same because of the gauge symmetry inherent
to the AOP.
This way we are allowed to dispense with 
the cumbersome converted fields and use
the corresponding 
Hopf term obtained in the AOP as the formula that truely 
represents the Hopf number to any configurations.
With this boundary condition we present
in sect.3 the Hamiltonian
formulation of the NSM in terms of the AOP and thereby
furnish a basis to quantize the model canonically.
We then examine in sect.4 
the angular momentum of the system
based on the canonical quantization scheme.
We shall find that the well-known formula [\BKW, \F] for
the fractional spin, 
$$
J_{\rm fractional} = \frac{\theta}{2 \pi}Q^{2}, 
\eqn\qsf
$$
holds for a specific class of configurations
(including the soliton {\it solutions}) but not
for generic configurations.
In sect.5 we construct soliton
operators explicitly in the (physical) Hilbert space.
This is just the AOP version of the soliton operators
previously proposed [\K], but it has an advantage in that
the construction is more transparent and that
a single operator can create a soliton
state whereas
in the previous construction 
we needed two operators defined in different patches on
the space $S^2$.  
Sect.6 is devoted to our conclusions and discussions.

\bigskip
\secno=2 \meqno=1 
\noindent
{\bf 2. Hopf Term}
\medskip

We wish to show in this section that it is 
possible (without using the conversion procedure)
to define the Hopf term that gives 
the Hopf integer to any configuration.  For this, 
we first introduce the AOP for the NSM.

The adjoint orbit of a group $G$ passing through an
element $K \in {\frak g}$, where ${\frak g}$ is 
the Lie algebra 
of the group $G$, is the subspace of ${\frak g}$ formed under
the adjoint action,
${\cal O}_K := \{\, gKg^{-1} \,
\vert \, g \in G \, \}$.
The orbit ${\cal O}_K$ 
is isomorphic to the coset space $G/H$ where $H$ is the 
isotropy group of the action at $K$.  For the $O(3)$ NSM 
we regard
the target space $S^2$ of the map $\nvec(x)$ as the coset
$SU(2)/U(1)$ which is obtained as the adjoint orbit
of $SU(2)$ passing through, say, $K = T_3$, where  
$\{T_a; a = 1, 2, 3\}$ is a basis of 
${\frak g} = {\frak su}(2)$.
In terms of the AOP our spin vectors read
\note
{
Convention: The basis of ${\frak su}(2)$ is taken to be 
in the defining representation 
$T_a = {{\sigma_a}\over{2i}}$, and our trace,
\lq $\tr := (-2)$ times the matrix trace\rq, is normalized 
as $\tr(T_aT_b) = \delta_{ab}$.  The epsilon
symbol $\epsilon_{\mu\nu\lambda}$ has the sign
$\epsilon_{012} = + 1$ and we use 
$\epsilon_{ij} := \epsilon_{0ij}$.   
}
$$
n(x) := n_1(x)T_1 + n_2(x)T_2 + n_3(x)T_3 
= g(x)\, T_3\, g^{-1}(x).
\eqn\aop
$$
Note that the constraint 
satisfied by the spin vectors, 
$\tr n^2(x) = 1$, 
is automatically fulfilled by the parametrization (\aop).
The AOP possesses redundancy  
with respect to the isotropy group $H$, which in our case
is the $U(1)$ group generated by the element $T_3$, as can be seen
from the fact that 
the same $\nvec(x)$ can be 
represented by different $G$-valued fields related by 
\lq gauge transformations\rq, 
$$
g(x) \longrightarrow g(x)\, h(x), \qquad 
\hbox{where}\quad h(x) \in H.
\eqn\gaugetr
$$ 

An important point to note [\PP, \KTT] 
is that the field $g(x)$ 
satisfying (\aop) becomes singular unless the corresponding
spin vector $\nvec(x)$ belongs to the zero soliton sector, as we
shall see shortly.
To deal with $g(x)$ everywhere regular,
we shall consider
a two dimensional disc $D^2$ whose boundary $\pa D^2$ 
is identified
with spatial infinity which is now the 
south-pole of the sphere $S^2$ (for the details, see [\KTT]).
Then, we shall define $g(x)$ as a map
$M := D^2 \times [0, T] \rightarrow SU(2)$.
With this definition we see that, 
if we let $g(\infty) \in SU(2)$ be 
some element satisfying 
$n(\infty) = g(\infty)\, T_3\, g^{-1}(\infty)$,
the boundary condition (\bdcon) is fulfilled if
$$
g(x) = g(\infty)\, k(x) 
\qquad \hbox{for} \quad x \in \pa D^2 \times [0, T],
\eqn\gbdcon
$$
with $k(x) = e^{\xi(x)T_3} \in H$ being  
an arbitrary smooth 
function over the boundary.
Let us observe that 
the AOP brings the soliton number to the form,
$$
Q(g) =  - {1\over{4\pi}} \int_{\pa D^2}\, \tr T_3 
(g^{-1}(x)dg(x)).
\eqn\soliton
$$
We then obtain $Q =  - {1\over{4\pi}}\int_{\pa D^2}\,d\xi(x)$,
which shows that the soliton
number is nothing but the winding number of the field $g(x)$ at
the boundary.  Note that gauge transformations (\gaugetr)
cannot change the soliton number because the function $h(x)$,
being defined over the contractible space $D^2$,
reduces to a trivial map
$\pa D^2 \simeq S^1 \rightarrow S^1$ 
at the boundary.
It is now clear that,
unless $Q(g) = 0$, the field $g(x)$ cannot be 
regular upon $S^2 \times [0, T]$, since 
the identification
$D^2$ with $S^2$ gives rise to a singularity at 
the south-pole for $Q(g) \ne 0$.  

Turning our attention to the Hopf term, we first recall
that in terms of the AOP the Hopf term (\quasihopf) reads
[\BMSS, \KTT]
$$
H(g) = {1\over{48\pi^2}} \int_{M}\, \tr (g^{-1}(x) dg(x))^3.
\eqn\hopf
$$
This term gives the Hopf integer for those $g(x)$ 
which are ${\bf x}$-independent at the both ends
$t = 0$ and $t = T$, since in this case
$M$ can be deformed to $S^3$ (by suspension) 
leading to
the familiar formula of the 
degree of mapping $S^3 \rightarrow SU(2) \simeq S^3$ 
as expected from (\pione).  To assign the Hopf integer 
to configurations having
non-vanishing soliton numbers, we consider those $g(x)$
satisfying the periodic condition in time up to a 
constant $h_{\rm c} \in H$,
$$
g({\bf x}, T) = g({\bf x}, 0)\, h_{\rm c}.
\eqn\periodic
$$
As discussed in our previous paper [\KTT], we may
assign the Hopf integer to this $g$ 
by using the formula (\hopf) with $g$ replaced by
$$
\bar g(x) := g(\infty)\, 
g^{-1}({\bf x}, 0) \, g({\bf x}, t).
\eqn\conv
$$
Indeed, this converted field $\bar g(x)$ 
becomes constant at the both ends of the period $[0, T]$,
and therefore $H(\bar g)$ can be used to provide 
the Hopf number to any $g$. 
(The factor $g(\infty)$
is inserted in (\conv) so that $\bar g(x)$ still
satisfies the same boundary condition as $g(x)$.)

But the price we pay for the use of the
Hopf term with the converted field 
is that, because of the
extra prefactors in (\conv), the Hopf term makes
the conventional treatise of the model, such as 
the Hamiltonian formulation, quite cumbersome.
To find a solution of the problem, 
let us compare the two formulae, $H(g)$ and $H(\bar g)$,
and see if the difference disappears under some condition.
The difference can be written as
$$
H(\bar g) - H(g) = Q(g)\,P(g),
\eqn\diff
$$
where $Q(g)$ is the soliton charge (\soliton) whereas
$$
P(g) :=  - {1\over{4\pi}} \int_0^T \, \tr T_3 
(g^{-1}(x)dg(x))\bigr\vert_{\pa D^2},
\eqn\timewind
$$
whose integration is along $t$ at fixed ${\bf x}$ on the boundary,
counts the winding number of the map $k(x)$ during the
period $[0, T]$.  More explicitly, it is given by
$P =  - {1\over{4\pi}}\int_0^T d\xi(x)$, which 
is independent of ${\bf x}$ on account of (\periodic)
and the smoothness of $k(x)$.  Note that $H(g)$ and $P(g)$
are not gauge invariant in general, 
in contrast to $H(\bar g)$ which is
gauge invariant.

We then notice that, 
if $k(x)$ is time-independent at the boundary $\pa D^2$, then
it follows that $P(g) = 0$ and hence
$H(\bar g) = H(g)$, implying that the conversion procedure
becomes unnecessary.  
Thus all we need to do is to render the boundary condition
(\gbdcon) slightly more restrictive, by
insisting that the function $k(x)$ be time-independent,
$$
k(x) = k({\bf x}) \qquad \hbox{on} \quad \pa D^2.
\eqn\newbdc
$$
In fact, this is not a real 
restriction for the boundary condition,
since the function $k(x)$ can always be made time-independent
by a gauge transformation (\gaugetr) with 
$h({\bf x},t) = k^{-1}({\bf x}, t)\, k({\bf x}, 0)$.
We therefore conclude that, under the boundary condition
(\gbdcon) with time-independent $k({\bf x})$, we can use
the same formula (\hopf) for the Hopf term to bestow 
a generic configuration $g(x)$ with the Hopf integer, 
even though the spacetime $M$ may not be deformable to $S^3$
without rendering the configuration singular.  
It should be stressed that,
without the periodicity
(\periodic) and the above mentioned boundary condition,
the Hopf term (\hopf) may not be
an integer nor gauge invariant,\note
{
Alternatively, one may define an integer-valued 
Hopf term  
by imposing, instead of (\newbdc), 
the strict periodicity $h_{\rm c} = 1$ in (\periodic).   
This however leads to a formula which is not 
gauge invariant due to $P(g)$.
} 
and that this is also true for the
conventional expression (\quasihopf).  We also mention that
the Hopf term $H(g)$ is gauge invariant in so far as 
the gauge transformation preserves the above boundary condition. 
The $O(3)$ NSM in the AOP hence possesses the residual 
gauge symmetry respecting the boundary condition
in the presence of the Hopf term.

\bigskip
\secno=3 \meqno=1
\noindent
{\bf 3. Hamiltonian Formulation}
\medskip

Having introduced the AOP and thereby defined the Hopf term for 
the $O(3)$ NSM, we now move on to furnish the 
Hamiltonian formulation of the model.  To this end, we first
note that in the AOP the action (\naction), supplemented with 
the Hopf term (\hopf), turns out to be   
$$
I = {1\over{2\lambda^2}}\int_{M} d^3 x \,
 \tr \bigl( g^{-1}(x) \pa_\mu g(x)\bigr\vert_{\frak r} \bigr)^2
+ {{\theta}\over{48\pi^2}} \int_{M}\, 
  \tr \bigl( g^{-1}(x) dg(x) \bigr)^3.
\eqn\gaction
$$
Here the symbol $X\bigr\vert_{\frak r}$ denotes 
the projected part of $X \in {\frak g}$ 
in the decomposition $X = X\vert_{\frak h}
+ X\vert_{\frak r}$ which is performed
according to the decomposition 
${\frak g} = {\frak h} \oplus {\frak r}$ where
${\frak r}$ is the orthogonal complement of the Lie subalgebra
${\frak h}$ of $H$.  In the present case
we have ${\frak r} = \hbox{span}\{T_1, T_2\}$ and hence, {\it e.g.},
the kinetic term in the action reads 
$\tr (g^{-1} \pa_\mu g\bigr\vert_{\frak r})^2
= (g^{-1} \pa_\mu g)_{a'}^2$, where 
the primed indices indicate
the components $a' = 1$, $2$ in the space ${\frak r}$. 
This, of course, is due to the fact that the target space of the
$O(3)$ NSM is $S^2$ rather than $SU(2)$.
In what follows, however, we introduce the canonical structure
of the NSM by symmetric reduction 
from that of the model defined on the $SU(2)$ group manifold
({\it i.e.}, the $SU(2)$ principal chiral model)
regarding the NSM as a constrained system.

To this end, let us introduce a set
of local coordinates $\{q^a; a = 1, 2, 3\}$ to parametrize
$g = g(q)$ and define the matrix $N_{ab}(q) 
:= (g^{-1}{{\pa}\over{\pa q^a}}g)_b(q)$ which is 
invertible [\BMSS].  From the Lagrangian 
in $I = \int d^3x\,{\cal L}$ we find
the canonical momentum conjugate to $q^a$,
$$
\pi_a :=\, {{\pa {\cal L}}\over{\pa(\pa_0 q^a)}} 
       =\, {1\over{\lambda^2}} N_{ab'}(g^{-1} \pa_0 g)_{b'}
+ {{\theta}\over{32\pi^2}}\,\epsilon_{ij}\, \epsilon_{bcd}\,
N_{ab}(g^{-1} \pa_i g)_{c}(g^{-1} \pa_j g)_{d},
\eqn\qmomentum
$$
which is supposed to satisfy the Poisson bracket relation,
$$
\bigl\{ q^a({\bf x})\, , \, \pi_b({\bf y})\bigr\} 
= \delta^b_a\,\delta({\bf x} - {\bf y}).
\eqn\canpm
$$

A much more convenient quantity than the
canonical momentum is
the `right-current' $R = R_a T_a$ with
$$
R_a := - (N^{-1})_{ab}\pi_b 
     = - {1\over{\lambda^2}} \delta_{ab'}(g^{-1} \pa_0 g)_{b'}
 - {{\theta}\over{32\pi^2}}\, \epsilon_{ij}\, \epsilon_{acd}\,
(g^{-1} \pa_i g)_{c}(g^{-1} \pa_j g)_{d}.
\eqn\rightcrt
$$
These components fulfill the relations,
$$
\bigl\{ R_a({\bf x})\, , \, R_b({\bf y})\bigr\} 
= \epsilon_{abc}\, R_c({\bf x})\,\delta({\bf x} - {\bf y}),
\qquad
\bigl\{ R_a({\bf x})\, , \, g({\bf y})\bigr\} 
= g({\bf x})\,T_a\,\delta({\bf x} - {\bf y}),
\eqn\rpb
$$
where $g$ is assumed to be in the defining representation.
Together with
$$
\bigl\{ g({\bf x})\, , \, g({\bf y})\bigr\} = 0,
\eqn\gpb
$$
the relations (\rpb) form the fundamental Poisson bracket
of the NSM in the AOP description at the non-reduced
level.  It is also worth mentioning that we can 
construct the `left-current',
$$
L := - g\,R\,g^{-1},
\eqn\leftcrt
$$
whose components $L = L_a T_a$ satisfy
$$
\bigl\{ L_a({\bf x})\, , \, L_b({\bf y})\bigr\} 
= \epsilon_{abc}\, L_c({\bf x})\,\delta({\bf x} - {\bf y}),
\qquad
\bigl\{ L_a({\bf x})\, , \, g({\bf y})\bigr\} 
= - T_a\, g({\bf x})\,\delta({\bf x} - {\bf y}).
\eqn\lpb
$$
The bracket relations, (\rpb) and (\lpb), show that
$R_a$ and $L_a$ are the generators of the right and
left actions on $g$, respectively, and hence they commute,
$$
\bigl\{ R_a({\bf x})\, , \, L_b({\bf y})\bigr\} = 0.
\eqn\commute
$$

The canonical structure of the $O(3)$ NSM
is then found by taking into account the constraint, 
$$
\Phi := R_3 + {{\theta}\over{32\pi^2}}\,
\epsilon_{ij}\,\epsilon_{c'd'}
(g^{-1} \pa_i g)_{c'}(g^{-1} \pa_j g)_{d'} = 0,
\eqn\constr
$$
derived from (\rightcrt).  To see if any secondary constraints
arise from this primary constraint, we consider the 
Hamiltonian,
$$
\eqalign{
{\cal H} :=&\, \pi_a \pa_0 q^a - {\cal L} \cr
          =&\, {{\lambda^2}\over{2}} \Bigl[ \, R_{a'} + 
{{\theta}\over{16\pi^2}}\,\epsilon_{ij}\,\epsilon_{a'b'}\,
(g^{-1} \pa_i g)_{b'}(g^{-1} \pa_j g)_{3} \, \Bigr]^2
 + {1\over{2\lambda^2}} (g^{-1} \pa_i g)_{a'}^2,
}
\eqn\hamiltonian
$$
where we have used the constraint (\constr) in the second line.
Then, using the relation
$$
\bigl\{ R_a({\bf x})\, , \, 
(g^{-1} \pa_i g)_{b}({\bf y}) \bigr\} 
= \epsilon_{abc}\, (g^{-1} \pa_i g)_{c}({\bf x})\,
\delta({\bf x} - {\bf y}) 
- \delta_{ab}\,\pa_i\delta({\bf x} - {\bf y})
\eqn\rcuraction
$$
derived from the fundamental Poisson bracket,
we can readily confirm that the constraint
(\constr) persists (strongly) in time,
$$
\bigl\{ \Phi({\bf x})\, , \, {\cal H}({\bf y})\bigr\} = 0,
\eqn\persist
$$
and hence no further constraints appear.
Combined with the involutive property,
$$
\bigl\{ \Phi({\bf x})\, , \, \Phi({\bf y})\bigr\} = 0,
\eqn\involution
$$
the persistency (\persist) implies that the constraint (\constr)
is first-class, and therefore generates a gauge symmetry ---
indeed it is the generator for 
the gauge transformation (\gaugetr).
It should be pointed out, however, that 
due to the second term in the constraint (\rightcrt)
the gauge transformation has a 
$\theta$-dependence, which shows up in the 
transformation of the quantities involving the current.

Models with first-class constraints are often
dealt with using the approach 
in which one finds a set 
of second-class constraints that contains the original first-class 
ones and thereby defines the Dirac bracket from the Poisson bracket so 
as to form the true Poisson bracket of the reduced system.  In the 
present paper we do not adopt this approach, and instead employ 
another approach where one considers gauge invariant quantities 
(physical observables) among 
which the Poisson bracket agrees with the 
Dirac bracket.  
Since the distinction between the \lq strong equality' 
and the \lq weak equality' is unnecessary 
in this approach, we use the 
ordinary equality symbol `$=$' throughout this paper, even when the 
weak equality symbol `$\approx$' is used in the former approach.

In this respect, we mention
a physically important observable, namely, 
the \lq magnetic field',
$$
B = \epsilon_{ij}\,\pa_i A_j,
\qquad \hbox{where} \quad A_j := (g^{-1}\pa_j g)_3,
\eqn\curvature
$$ 
which is 
gauge invariant $\{ \Phi\, , \, B \} = 0$ and gives rise to
the flux penetrating the disc $D^2$ proportional to 
the soliton charge, 
$$
\int_{D^2}d{\bf x}\, B =  - 4\pi Q.
\eqn\bsoliton
$$  
The magnetic field also appears in the
constraint (\constr) which can be rewritten as
$\Phi = R_3 - {\theta\over{16\pi^2}}B$.   
More generally, let us consider
the fields,
$$
B_a := \epsilon_{ij}\,\pa_i(g^{-1}\pa_j g)_a,
\eqn\magnet
$$
and construct the current ${\cal R} = {\cal R}_a T_a$
with
$$
{\cal R}_a := R_a - {\theta\over{16\pi^2}}B_a.
\eqn\cright
$$
Note that the constraint (\constr) is now the third
component of the new right-current,
$$
{\cal R}_3 = \Phi = 0.
\eqn\newconstr
$$
Observe also that the current is 
covariant under the gauge transformation,
$$
\bigl\{ \Phi({\bf x})\, , \, {\cal R}_{a'}({\bf y})\bigr\} 
= \epsilon_{a'b'}\, {\cal R}_{b'}({\bf x})\,
\delta({\bf x} - {\bf y}).
\eqn\covtr
$$
In fact, this is part of the salient property that 
the convariant right-current (\cright) forms exactly
the same Poisson bracket\note{
Equivalently, one can also say that,
although one finds a 
functional potential in (\cright) which gives rise to a holonomy
in the functional space, 
the corresponding functional curvature
vanishes [\WZb, \PP].}
as the original right-current,
$$
\bigl\{ {\cal R}_a({\bf x})\, , \, {\cal R}_b({\bf y})\bigr\} 
= \epsilon_{abc}\, {\cal R}_c({\bf x})\,\delta({\bf x} - {\bf y}),
\qquad
\bigl\{ {\cal R}_a({\bf x})\, , \, g({\bf y})\bigr\} 
= g({\bf x})\,T_a\,\delta({\bf x} - {\bf y}).
\eqn\rpb
$$
Being both covariant and fundamental (under the Poisson bracket),
the current components $ {\cal R}_{a'} $, together with 
the components $(g^{-1}\pa_i g)_{a'}$,
can be used as basic building blocks for
constructing physical observables.  

{}For instance, since ${\cal R}_{a'} = - {1\over{\lambda^2}}
(g^{-1} \pa^0 g)_{a'}$ 
the symmetric energy-momentum tensor
derived from the action (\gaction),
$$
T^{\mu\nu} = {1\over{\lambda^2}} 
(g^{-1} \pa^\mu g)_{a'}(g^{-1} \pa^\nu g)_{a'}
- {1\over{2\lambda^2}}\, \eta^{\mu\nu}\, 
(g^{-1} \pa^\rho g)_{a'}^2,
\eqn\emtensor
$$
is found to be built up only with those covariant elements
and is hence manifestly gauge invariant.
In particular, 
the components for energy and momentum read
$$
T^{00} = {{\lambda^2}\over{2}}\, {\cal R}_{a'}^2
 + {1\over{2\lambda^2}}\, (g^{-1} \pa_i g)_{a'}^2 =  {\cal H},
\eqn\etenergy
$$
and
$$ 
T^{0i} = {\cal R}_{a'}\, (g^{-1} \pa_i g)_{a'},
\eqn\etmom
$$
respectively.

With respect to the covariant right-current ${\cal R}$ we can 
define the new left-current, 
$$
{\cal L} := - g\,{\cal R}\,g^{-1} = L + {\theta\over{16\pi^2}}
\,\epsilon_{ij}\,\pa_i(\pa_j g\,g^{-1}).
\eqn\newlftcrt
$$
Clearly, the components of 
the current ${\cal L} = {\cal L}_a T_a$ satisfy the
same relations as before,
$$
\bigl\{ {\cal L}_a({\bf x})\, , \, {\cal L}_b({\bf y})\bigr\} 
= \epsilon_{abc}\, {\cal L}_c({\bf x})\,\delta({\bf x} - {\bf y}),
\qquad
\bigl\{ {\cal L}_a({\bf x})\, , \, g({\bf y})\bigr\} 
= - T_a\, g({\bf x})\,\delta({\bf x} - {\bf y}),
\eqn\newlpb
$$
and commute with the covariant right-current,
$$
\bigl\{ {\cal R}_a({\bf x})\, , \, {\cal L}_b({\bf y})\bigr\} = 0.
\eqn\newcommute
$$
Being gauge invariant, all the components ${\cal L}_a$ are  
physical observables.  Note that,
as a vector, the physical left-current is orthogonal
to the spin vector,
$$
{\cal L}_a n_a = \tr ({\cal L}\,n)
= \tr (g^{-1}{\cal L}g)T_3 = - {\cal R}_3 = 0,
\eqn\ortho
$$
on account of the constraint (\newconstr).

To complete our gauge invariant approach, 
it is necessary to find a physical
observable corresponding to the field $g$.  
One obvious way to do this is to go back to
the spin vector ${\bf n}(x)$.  The defining relation
(\aop) of the AOP shows that it is trivially gauge invariant, 
$
\bigl\{ \Phi({\bf x})\, , \, n_a({\bf y}) \bigr\} = 0,
$
and that it transforms as isovector with respect to the
left-action,
$$
\bigl\{ {\cal L}_a({\bf x})\, , \, n_b({\bf y})\bigr\} 
= \epsilon_{abc}\, n_c({\bf x})\,\delta({\bf x} - {\bf y}).
\eqn\nisovec
$$

Based on the Hamiltonian formulation of the NSM developed
above, we carry out the canonical quantization
by replacing the Poisson bracket with the commutator,
$\{\,\,\, ,\,\,\} \rightarrow {1\over i}[\,\,\, ,\,\,]$, after
promoting the classical quantities $A$ to 
the corresponding quantum operators $\widehat A$.  We shall
use the quantum language in sect.5 to construct soliton
operators explicitly.  Prior to this, however, 
in the next section we wish to
examine the issue of fractional spin which 
has been argued to occur in the presence of the Hopf term.

\bigskip
\secno=4 \meqno=1
\noindent
{\bf 4. Fractional Spin}
\medskip

In this section, following the line of argument
given in [\BKW],  
we shall study the angular momentum of the
system paying particular attention to 
the $\theta$-dependence of 
the angular momentum which is the source of fractional spin.
In $(2+1)$-dimensions, the space rotation group $SO(2)$ is 
Abelian and the generator is ambiguous up to the addition
of a constant.  This ambiguity can be removed [\BKW] by 
embedding the $SO(2)$ in the full \lq Lorentz group' 
$SO(2,1)$ and thereby define the angular momentum as
$$
J := \int d{\bf{x}}\, \epsilon_{ij}\,x^{i}\,T^{0j}
   = \int d{\bf{x}}\, \epsilon_{ij}\,x^{i}\,{\cal R}_{a'}\,
(g^{-1}\partial_{j}g)_{a'}.  
\eqn\angm
$$
The angular momentum is a physical observable since its
gauge invariance,
$\{ \Phi({\bf x})\, , \, J \}=0$,
follows from the invariance of $T^{0j}$.
It acts as a generator for the infinitesimal spatial rotation, 
$$
\bigl\{ J \, , \, g \bigr\}
= \epsilon_{ij}\,x^i \partial_j \,g
-\epsilon_{ij}\,x^i \,(g^{-1}\partial_jg)^3 g \, T_3,
\eqn\rotg
$$
leading to the transformation of the spin vector
as scalar,
$\bigl\{ J \, , \, n_a \bigr\}
=\epsilon_{ij}\,x^{i}\, \partial_{j} n_a$.
We note in passing that in the application for 
condensed matter systems where, {\it e.g.},
${\bf n}(x)$ represents 
the spin of the valence electrons, the field ${\bf n}(x)$ 
transforms as a vector 
under the rotation about the axis perpendicular
to the plane.  In our case, however, 
we assume ${\bf n}(x)$ to be a scalar for the reason that 
we focus on the possible fractional 
orbital angular momentum which arises in the
$\theta$-dependent part in $J$.

In order to discuss the $\theta$-dependence of 
the angular momentum, let us split it into two parts
$J = J_1 + J_2$ with
$$
J_1 :=\int d{\bf{x}}\,
\epsilon_{ij}\,x^i\,R_{a'}\,(g^{-1}\partial_{j}g)_{a'}, 
\qquad
J_2 :=-\frac{\theta}{16 \pi^{2}}\int d{\bf{x}}\,
\epsilon_{ij}\,x^i\,B_{a'}\,(g^{-1}\partial_{j}g)_{a'}.
\eqn\spinpart
$$
If we use 
$B_{a'}=-\epsilon_{kl}\epsilon_{a'b'}
(g^{-1}\partial_{k}g)_{b'}(g^{-1}\partial_{l}g)_{3}$ 
obtained from (\magnet), we can rewrite the 
manifestly $\theta$-dependent part $J_2$ purely in terms of
the gauge field $A_{i}$ and its curvature $B$ 
in (\curvature) as
$$
\eqalign{
J_2 &= \frac{\theta}{16 \pi^{2}}\int \,d\hbox{\bf x} \,
 \epsilon_{ij}\,x^{i}\,\epsilon^{kl}
(g^{-1}\partial_{k}g)^{3}\,
\tr \bigl( T_{3}\,[\, g^{-1}\partial_{l}g 
\,,\,g^{-1}\partial_{j}g\,] \bigr) \cr
      &= \frac{\theta}{16 \pi^{2}}\int d\hbox{\bf x}
\,\epsilon_{ij}\,x^{i} A_{j} B.
}
\eqn\jtwo
$$
We stress that, in general,
neither $J_1$ nor $J_2$ is gauge invariant in itself,
$$
\bigl\{\Phi({\bf x})\, ,\, J_{1}\bigr\}
= - \bigl\{\Phi({\bf x})\, ,\, J_{2}\bigr\}
=\epsilon_{ij}\,x^i\,\partial_{j} B({\bf x}).
\eqn\gitotal
$$
The exception occurs when the curvature depends only
on the radial distance $r = \vert{\bf x}\vert$ 
of the disc $B({\bf x}) = B(r)$, in which case  
$J_1$ and $J_2$ are separately gauge invariant.  

It is interesting to observe that
the part $J_2$ admits a concise formula
in the Coulomb gauge $\partial_{i}A_{i}=0$,
$$
J_2=\frac{\theta}{4 \pi}Q^{2}, 
\eqn\sqsoliton
$$
allowing for the interpretation that 
the fractional ($\theta$-dependent) part is 
proportional to the square of the soliton number
of the configuration under consideration.  
The derivation of the $Q^2$-formula (\sqsoliton)
is essentially the same
as the one which has been given 
for evaluating the similar (but not exactly
the same; see below) part in the angular momentum [\BKW].
Namely, we first write the potential as
$A_{i}=\epsilon_{ij}\partial_{j}\rho$, and 
find that the function
$\rho$ is given by
$\rho({\bf x})=\int d{\bf x}\,
D({\bf x}-{\bf y})\,B({\bf y})$ 
with
$D({\bf x}-{\bf y}) =-\frac{1}{\Delta}
\delta ({\bf x}-{\bf y}) 
=-\frac{1}{4\pi}\ln |{\bf x}-{\bf y}|^{2} $
being the inverse of the Laplacian 
$\Delta = \partial_{i}^{2}$.
We then have 
$$
A_{i}({\bf x})=\epsilon_{ij}\,\partial_{j}
\int d{\bf y}\,\Bigl( -\frac{1}{4\pi}\ln |{\bf x}-{\bf y}|^{2}
\,B({\bf y})\Bigr).
\eqn\solpotent
$$
Substituting (\solpotent) into (\jtwo) we get
$$
\eqalign{
J_2&=\frac{\theta}{16 \pi^{2}}
\Bigl(-\frac{1}{4\pi}\Bigr) 
\int d{\bf x}\, d{\bf y}\,\epsilon_{ij}\,x^i\,
\epsilon_{jk}B({\bf x})
\left( \partial_{k} \ln |{\bf x}-{\bf y}|^{2} \right) 
B({\bf y}) \cr
&=\frac{\theta}{32 \pi^{3}}
\int d{\bf x}\, d{\bf y}
\frac{x_{i}\,(x_{i}-y_{i})}{|{\bf x}-{\bf y}|^{2}}
B({\bf x})\,B({\bf y}), 
}
\eqn\evajtwo
$$
which proves (\sqsoliton) on account of (\bsoliton).

We cannot, however, conclude from this result that the
fractional spin of the system always occurs according to
the square charge rule (\sqsoliton), simply because
the value of $J_2$ depends in general 
on the gauge chosen; in other words,
$J_2$ is not a physical observable for a generic
configuration.  Moreover, as we shall see
more explicitly soon, the part $J_1$ also possesses an implicit 
$\theta$-dependence in view of the relation (\rightcrt).  
Clearly, the correct $\theta$-dependence of the spin,
or the physically meaningful value of the fractional spin,
can be obtained only when one succeeds to separate the gauge
invariant $\theta$-dependent part from the total $J$.
This however
seems impossible if one is to use
the set of covariant/invariant 
elements constructed in sect.3  
out of the basic variables in the phase space, and has
certainly been unaccomplished by anyone so far.  Flawed with 
this problem, 
the $Q^2$-formula (\qsf) for the
fractional part of angular momentum 
obtained earlier [\BKW, \PP, \F]
(where the factor of the formula is different from ours 
due to the different split employed) is untenable as
a formula for a generic configuration.

Let us now consider a class of 
specific configurations to
illustrate the gauge 
dependence in the split of the angular momentum and to 
examine how the angular momentum can be fractionalized 
through the physical $\theta$-dependence.  
To this end, we introduce
the polar coordinates 
${\bf x}=(r \cos\varphi, \, r \sin\varphi)$ 
with $(r,\, \varphi)\,\in [0,1] \times [0,2\pi]$ to
parametrize the disc $D^{2}$ of unit radius.
We then take the configuration,
$$
{\bf n}({\bf x},t)=
\Bigl( \cos \bigl(\alpha({\bf x})+ \phi(t)\bigr)\, 
\sin\beta({\bf x})\, ,\,
\sin \bigl(\alpha({\bf x})+ \phi(t)\bigr)\, 
\sin\beta({\bf x})\, ,\,
\cos\beta({\bf x}) \Bigr),
\eqn\confn
$$
where $\alpha(r,\,\varphi)$ and $\beta(r,\,\varphi)$ 
are time independent functions representing  
a generic static configuration 
possessing the soliton number $n$, say.
The only dynamical variable $\phi(t)$, on the 
other hand, is the collective coordinate representing, say, $m$-times 
the revolution of the static configuration around the origin ${\bf x} 
= 0$ during the time period $[0, T]$.  These topological requirements 
will be met if we assume the boundary conditions,
$$
\alpha(r, 2\pi) - \alpha(r, 0) = 2n\pi, \qquad 
\beta (1, \varphi)=\pi, \quad \beta (0,\varphi)=0, \qquad 
\phi(T) - \phi(0) = 2m\pi.
\eqn\bdcvbs
$$ 
The spin vector field (\confn) is realized in the AOP by 
$$
g({\bf x},t) = e^{(\alpha({\bf x}) + \phi(t)) T_3}\, 
e^{\beta({\bf x}) T_2}\, 
e^{- (\alpha({\bf x}) - \phi(t)) T_3} \, 
e^{\eta({\bf x},t) T_3}, 
\eqn\cfg
$$
where we have introduced the 
function $\eta({\bf x}, t)$ (which is regular
over $D^2$) to account for the gauge freedom.  
This configuration satisfies the condition (\newbdc) if 
$\eta({\bf x}, t)$ is 
time independent ${d\over{dt}} \eta({\bf x}, t)= 0$
on the boundary $\pa D^2$.  With the above boundary
conditions (\bdcvbs) one can easily confirm that
our configuration (\cfg) has the following
soliton and Hopf numbers,
$$
\eqalign{
Q(g)&=-\frac{1}{4 \pi}\int_{\partial D^{2}} 
d\alpha \, (\cos \beta-1) \, =
\frac{1}{4 \pi} \times 2n \pi \times 2=n,\cr
H(g)&=\frac{Q(g)}{2 \pi} \int_0^T d\phi
= \frac{1}{2 \pi} \times n \times 2m \pi = nm.
}
\eqn\shnumbers
$$

Under the class of configurations (\cfg) we have
the potential 
$A_{i} = \pa_i\alpha(\cos\beta - 1)+\partial_{i}\eta$
and hence the curvature,
$$
B = \frac{1}{r}
 \Bigl( \partial_{\varphi}( \cos \beta )\, \partial_{r}\alpha-
 \partial_{\varphi} \alpha\, \partial_{r}(\cos \beta) \Bigr).
\eqn\curvsp
$$
The gauge dependence of the $J_2$ part can now be explicitly
demonstrated by taking, for instance, the configuration,
$\alpha(r,\varphi)=\frac{n}{2 \pi}\varphi^{2}$, 
$\beta(r,\varphi)=r$ for which we have 
$B=\frac{\sin (r)}{r}\frac{n \varphi}{\pi}$.  
Since the
curvature depends on $\varphi$, we learn from our 
earlier discussion that
$J_2$ cannot be gauge invariant.  In fact, substituting
the potential and the curvature in (\jtwo) we find
$J_{2}=\frac{\theta}{3 \pi} n^{2}$ for, {\it e.g.}, $\eta = 0$, 
which disagrees with the result (\sqsoliton) obtained
in the Coulomb gauge.  It even departs from the $Q^2$ rule
for $\eta = \varphi(\varphi-2\pi)$ where we get
$J_{2}=\frac{\theta}{3 \pi}n^{2}+\frac{\theta}{6}n$.

This disagreement stems, of course, from the fact
that our potential $A_i$ does not satisfy the Coulomb gauge.
However, behind this lies the important question as to how
the fractional part of the angular momentum can be 
defined gauge invariantly.  To examine this issue,
let us evaluate the part $J_1$ and see how the
gauge non-invariance disappears when combined with $J_2$.
Plugging the configuration (\cfg) in (\spinpart), we find
$$
J_{1}=- N \dot{\phi}
- \frac{\theta}{16 \pi^{2}}\int d{\bf x}\, 
\epsilon_{ij}\,x^{i}A_{j}B,
\eqn\spinone
$$
with $N = \frac{1}{\lambda ^{2}}
\int d{\bf x} \,\sin^{2}\beta\, \pa_{\varphi}\alpha$.  
To express $\dot \phi$ in terms of the
conjugate momentum $\Pi_\phi$, we obtain 
from the action $I = \int dt\, L$ in (\gaction) the Lagrangian
for the collective mode in (\cfg),
$$
L= - M + \frac{K}{2}\, \dot{\phi}^{2}
+ \frac{\theta}{2\pi}\,n\,\dot{\phi},
\eqn\lagphi
$$
where
$
M = \frac{1}{2\lambda^{2}}\int d{\bf x} 
\left(\sin^{2}\beta(\partial_{i}\alpha)^{2}
+(\partial_{i}\beta)^{2} \right)
$
is the mass of the static configuration and
$K = \frac{1}{\lambda ^{2}}
\int d{\bf x} \,\sin^{2}\beta$
is the `nonrelativistic mass' of the 
collective mode excitation.
The momentum is then obtained as
$$
\Pi_{\phi}=\frac{\partial  L}{\partial \dot{\phi}}
= K \dot{\phi}+\frac{\theta}{2 \pi}\,n.
\eqn\momentum
$$
Thus, with the `moment of inertia',
$$
I: = {N \over K} = 
\frac{\int d{\bf x} \,\sin^{2}\beta \,\partial_{\varphi}\alpha}
{\int d{\bf x} \,\sin^{2}\beta },
\eqn\inertia
$$ 
we arrive at the formula,
$$
J_{1} = - \left(\Pi_{\phi}-\frac{\theta}{2 \pi}\,n \right) I
- \frac{\theta}{16 \pi^{2}}\int d{\bf x}\, 
\epsilon_{ij}\,x^{i}A_{j}B.
\eqn\finalspinone
$$
It is now clear that the last term in (\finalspinone)
is gauge dependent but canceled precisely when combined with
$J_2$ in (\jtwo), leaving the first term as the physical
total angular momentum $J$ of the system.

Upon quantization, the dynamical 
variables $\phi$ and $\Pi_\phi$ are promoted to 
the operators $\widehat \phi$ and $\widehat{\Pi}_\phi$ satisfying
the commutator $[\widehat{\Pi}_{\phi},\,\widehat\phi]=\frac{1}{i}$. 
In the coordinate representation we have 
$\widehat{\Pi}_{\phi} = \frac{1}{i}\frac{d}{d\phi}$,
and therefore the wave functions $\Psi_k(\phi) = e^{ik\phi}$
for $k \in \Z$ provide the eigenfunctions of the Hamiltonian
obtained from the Lagrangian (\lagphi),
$$
\widehat H = M + {1 \over {2K}}
\left(\widehat{\Pi}_{\phi}-\frac{\theta}{2 \pi}\,n \right)^2,
\eqn\hamiltonian
$$
with the eigenvalues,
$$
E_k = M + {1 \over {2K}} 
\left(k - \frac{\theta}{2 \pi}\,n \right)^2.
\eqn\eigenenergy
$$
The eigenstates have the total angular momentum,  
$$
J_k = - \left(k - \frac{\theta}{2 \pi}\,n \right)I,
\eqn\totaleigenam
$$
which shows that the fractional spin part is 
independent of the energy level and given by
$$
J_{\rm fractional} = \frac{\theta}{2 \pi}\,n I.
\eqn\fracspin
$$

At this point we note that, because of the boundary condition 
(\gbdcon) with (\newbdc), the only global symmetry of 
the original action (\gaction)
is the one under the combined transformation
of the spatial $U(1)$ rotation, generated by letting
$\varphi \rightarrow \varphi + \epsilon$, 
and the isospin $U(1)$ rotation with respect to $T_3$,
generated by $g \rightarrow e^{\delta T_3} g$.
Thus we may restrict our attention to configurations
which exhibit this symmetry in the class (\cfg)
we are considering.  An obvious case in which 
this symmetry is realized occurs (with $\delta = - n \epsilon$)
when we have the functions $\alpha$ and $\beta$ of the form,
$$
\alpha (r, \varphi)=n\varphi, \qquad 
\beta (r, \varphi)=\beta(r).
\eqn\rotfunction
$$
It follows from (\curvsp) that in this case 
the magnetic field becomes 
rotationally invariant $B = B(r)$, and  
therefore both $J_1$ and $J_{2}$ are gauge invariant.
Thus the $Q^2$ rule (\sqsoliton) for $J_2$ indeed holds but,
as we have seen above,
this alone does not determine the $\theta$-dependence of
the angular momentum because $J_1$ is also   
$\theta$-dependent.  The correct formula for 
the fractional spin part in the present case can be obtained from
(\fracspin) by noticing that the moment of inertia (\inertia)
coincides with the soliton number 
$$
I = n,
\eqn\mieqsol
$$ 
for the configurations (\rotfunction), leading to
$$
J_{\rm fractional} = \frac{\theta}{2 \pi}\,n^2,
\eqn\rfracspin
$$
in agreement with (\qsf).  

Thus we conclude
that the $Q^2$-formula (\qsf) for fractional spin
is valid for configurations of the type (\rotfunction)
for which the curvature $B$ depends only on the radial direction.
The known $n$-soliton solution actually 
falls into this restricted
class, but generic ones having soliton number $n$ do not,
and for those the formula (\qsf) does not hold, as seen by 
(\fracspin) which is obtained for our
class (but not restricted one) 
of configurations
for which $I$ is not necessarily $n$.
We recall that in the literature one normally splits 
the angular momentum $J$ with the intention that the first
part $J_1$ becomes the ordinary orbital angular momentum 
implementing
the spatial $U(1)$ rotation for the spin vectors ${\bf n}(x)$
while the second part $J_2$ represents the additional
contribution induced by the Hopf term --- hence 
proportional to the angle parameter $\theta$ --- causing 
the fractional spin.  Our discussion in this section
shows that the angular momentum does not
seem to admit the split in the way intended; indeed, our
$J_1$ in (\spinpart) does meet the requirement and yet
is not $\theta$-independent nor a physical observable.

\bigskip
\secno=5 \meqno=1
\noindent
{\bf 5. Soliton Operators}
\medskip

The `soliton operator' which generates a soliton state has
previously been constructed in [\K] in order to discuss 
the fractional spin at the quantum level. 
The operator introduced there creates a single soliton 
on the `classical vacuum state', {\it i.e.}, the eigenstate
of the spin vector operator 
$\widehat{n}(x)$ 
with the constant eigenvalue $n(\infty)$ (for all $x$)
specified by the boundary condition (\bdcon).
Below we shall argue that in the AOP the construction of the operator
becomes more transparent and further that 
it can be generalized to operators
creating a state concentrated about a generic classical configuration.

To begin with, we note that the
Hilbert space of the NSM may be defined as the
physical space ${\cal H}_{\rm phys}$ 
consisting of states satisfying the physical state condition, 
$\widehat\Phi\, |{\rm phys} \rangle = 0$, in the entire 
Hilbert space ${\cal H}$ of
the unconstrained model, that is, the principal chiral model.
Being a physical observable, 
the spin vector operator given by 
$$
\widehat{n}(x) = \widehat{n}_a(x)\, T_a = 
\widehat{g}(x)\,T_3\,\widehat{g}^{-1}(x),
\eqn\spaop
$$
admits the coordinate representation within
the physical space ${\cal H}_{\rm phys}$: 
$$
\widehat{n}(x)\, |{n}(x) \rangle 
= {n}(x)\, |{n}(x) \rangle.
\eqn\estates
$$
Among the eigenstates is the classical vacuum state mentioned
above,
$$
\widehat{n}(x)\, |{n}(\infty) \rangle 
= {n}(\infty)\, |{n}(\infty) \rangle.
\eqn\cvs
$$

Now, suppose we are given two arbitrary 
classical configurations, ${n}_{1}(x)$ and ${n}_{2}(x)$, 
possessing different soliton numbers in general.
Then there are corresponding states in ${\cal H}_{\rm phys}$
having these configurations as eigenvalues.  What we shall show
below is that in the AOP the unitary 
operator which relates these two states,
$$
|{n}_2(x) \rangle = \widehat U \, |{n}_1(x) \rangle,
\eqn\uo
$$
can be constructed along the method of Ref.[\K] with more ease
and transparency.  To this end, let $g_1(x)$ and $g_2(x)$
be the fields associated with the two spin vector fields
under the AOP, namely,
${n}_{1} = g_1\, T_3\, g_1^{-1}$ and
${n}_{2} = g_2\, T_3\, g_2^{-1}$, respectively.
We then have the relation,
$$
n_{2}(x)= g_{21}(x)\,n_{1}(x)\, g_{21}^{-1}(x),
\eqn\relation
$$
with $g_{21} = g_2\, g_1^{-1}$.  Thus it is clear that
the transformation on the spin vector ${n}_1 
\rightarrow {n}_2$ can be achieved by 
the transformation $g_1 \rightarrow g_2 = g_{21}\, g_1$
in the AOP.
Since this is a left-action on the field, it
can be implemented if we take the unitary operator $\widehat U$ 
to be the representation $\widehat U(g_{21})$  
of the group action in the Hilbert space.

Explicitly, if we use, {\it e.g.},
the Euler angle decomposition for the element $g_{21}$,
$$
g_{21}(x)= e^{\alpha(x)T_3}\,e^{\beta(x)T_2}\,e^{\gamma(x)T_3},
\eqn\ead
$$
then the unitary operator implementing the left-action is given by 
$$
\SO= 
  e^{{1\over i}\int d{\bf x} \alpha(x){\cL}_{3}(x)}
  e^{{1\over i}\int d{\bf x} \beta(x){\cL}_{2}(x)}
  e^{{1\over i}\int d{\bf x} \gamma(x){\cL}_{3}(x)},
\eqn\sop
$$
where $\widehat{\cal L}_a$ are the operators corresponding to
the gauge invariant 
left-current components ${\cal L}_a$.  It then follows from their 
commutation relations implied by the Poisson bracket (\newlpb),
$$
\bigl[ \widehat{\cal L}_a({\bf x})\, , \, 
\widehat{\cal L}_b({\bf y}) \bigr] 
= i\, \epsilon_{abc}\, \widehat{\cal L}_c({\bf x})\,
\delta({\bf x} - {\bf y}),
\qquad
\bigl[ \widehat{\cal L}_a({\bf x})\, , \, \widehat g({\bf y}) \bigr] 
= - i\, T_a\, \widehat g({\bf x})\,\delta({\bf x} - {\bf y}),
\eqn\comnewlpb
$$
that the unitary operator (\sop) provides a representation,
$\widehat U(g_1)\,\widehat U(g_2) = \widehat U(g_1 g_2)$,
and that it gives the left-action,  
$$
\widehat{U}^{-1}(g_{21})\,\qg(x)\, \SO = g_{21}(x)\,\qg(x),
\eqn\conj
$$
as intended.  Using (\spaop) and (\conj) we obtain
$$
\eqalign
{
\widehat{n}(x) \,\SO\,|{n}_{1}(x) \rangle 
&= \SO\, g_{21}(x)\, \widehat{n}(x)\, g_{21}^{-1}(x)\,
   |{n}_{1}(x) \rangle \cr
&= g_{21}(x)\,n_{1}(x)\,g_{21}^{-1}(x)\,\SO\,|{n}_{1}(x)\rangle\cr
&= n_{2}(x)\, \SO\,|{n}_{1}(x) \rangle,
}
\eqn\crel
$$
which shows that 
the unitary operator $\SO$ does the job 
required in (\uo) and hence can be regarded as the operator
creating the state $|{n}_{2}(x) \rangle$ out of 
$|{n}_{1}(x) \rangle$.  In particular, when the state
$|{n}_{1}(x) \rangle$ is the classical vacuum state 
$|{n}(\infty) \rangle$ in (\cvs) and the newly created 
state $|{n}_{2}(x) \rangle$ has a nonvanishing soliton number,
the unitary operator $\widehat U$ may be thought of as 
a soliton operator.  Note that the unitary operator that has the
effect (\crel) is not unique, since the same
construction (\sop) using the operators corresponding to
the components of the original left-current 
$\widehat L_a$
yields the identical result (\crel).  However, in this case
the resultant state $\SO\,|{n}_{1}(x) \rangle$ 
cannot belong to the physical Hilbert space 
${\cal H}_{\rm phys}$, because unlike $\widehat {\cal L}_a$
the operators $\widehat L_a$ are not gauge invariant and do not
commute with $\widehat\Phi$.  We also note that, since
the unitary operator $\SO$ is defined with respect to the
eigenstates $|{n}(x) \rangle$ used for the 
coordinate representation (\estates), it
has an intrinsic ambiguity associated with the choice
of the eigenstates which are determined up to the form,
$\widehat U(e^{\xi(x)n(x)}) \,|{n}(x) \rangle$ with
$\xi(x)$ being a function.

{}Finally, we show that the unitary operator (\sop)
does possess the correct soliton number required to
create the state $|{n}_{2}(x) \rangle$ from $|{n}_{1}(x) \rangle$.
Indeed, from (\conj) we observe that the soliton
operator $\widehat Q(g) := Q(\widehat g)$ transforms as
$$
\widehat U^{-1}(g_{21})\, \widehat{Q}(g)\, \SO 
=-\frac{1}{4\pi}\int_{\partial D^{2}} 
\hbox{Tr}\,T_{3}(\qg^{-1} d \qg)
-\frac{1}{4\pi}\int_{\partial D^{2}}
\hbox{Tr}\,\widehat{n}\, (g^{-1}_{21} d g_{21}).
\eqn\scexchange
$$
Since the two configurations $g_1$ and $g_2$ share the same
boundary value $g(\infty)$ as 
$g_1(x) = g(\infty)k_1(x)$, $g_2(x) = g(\infty)k_2(x)$
with $k_1$, $k_2 \in H$, we have at spatial boundary,
$$
\hbox{Tr}\,\widehat{n}\, (g^{-1}_{21} d g_{21})\bigr\vert_{\pa D^2}
= \hbox{Tr}\,T_3(k_2 k_1^{-1})^{-1}d(k_2 k_1^{-1})
= \hbox{Tr}\,T_3(k_2^{-1} d k_2) - \hbox{Tr}\,T_3(d k_1\, k_1^{-1}).
\eqn\spbsol
$$
Hence (\scexchange) becomes
$$
\widehat U^{-1}(g_{21})\,\widehat{Q}(g)\,\SO
=\widehat{Q}(g) + Q(g_{2}) - Q(g_{1}),
\eqn\shift
$$
that is, the operator $\SO$ shifts the soliton number  
precisely by the amount of the soliton charge difference
between the two configurations, $g_1$ and $g_2$.
We point out that our formulation of the soliton operator
(\sop) does not require the patching procedure needed
in the previous construction [\K] where one needs two
operators defined on two different 
patches covering the space $S^2$.
The reason for this is basically due to the trivialization
of the space $S^2 \rightarrow D^2$, where we traded the
topological property of the spin field $n(x)$ for
the behavior of the field $g(x)$ at the boundary (\gbdcon).

\bigskip
\secno=5 \meqno=1
\noindent
{\bf 6. Conclusions and Discussions}
\medskip

We have seen in this paper that the AOP, which
employs $SU(2)$ group-valued fields $g(x)$ defined on
the space $D^2$ for describing the spin vectors of the NSM,
is useful in analyzing the topological aspect of the model. 
We first provided a boundary condition at spatial
infinity ({\it i.e.}, the boundary $\pa D^2$) that renders
the conversion procedure unnecessary, which was required
earlier to provide the Hopf term for generic configurations.
After this, we presented the Hamiltonian formulation
of the NSM in the AOP, where we 
observed that the model can be interpreted as 
a constrained system of the $SU(2)$ principal chiral model. 
The constraint is first-class and hence generates
a gauge symmetry that corresponds to the ambiguity in
the AOP.   Accordingly, as a reduced system the NSM is 
described by a full set of physical observables consisting
of gauge invariant quantities.  The gauge invariance
turned out to be crucial in sorting out the correct, physically
meaningful, fractional spin part in the total angular momentum 
of the model.  We found that the $Q^2$-formula for 
fractional spin proposed earlier 
does not hold in general, although it 
is correct for a restricted class
of configurations which includes the soliton solutions.
It should however be noted that the problem
of fractional spin can be addressed (and answered) properly
only at the quantum level, rather than the classical level [\CM],
and for this 
we seem to lack a method to analyze the problem on a general basis 
except that of 
using a collective coordinate about a specific class
of configurations as we did here. 
The gauge invariance was also used to determine the 
soliton operator which creates the state concentrated around
a generic classical configuration upon the classical
vacuum state.  Our construction is based on the observation
that the soliton operator is given by the 
unitary representation of the left-action in the Hilbert 
space implementing the transition from one configuration
to the other.  Being group-valued, the AOP is most appropriate
to this purpose.

Our AOP description of the NSM may be extended 
to the general coset nonlinear models over $G/H$
along the line of Ref.[\DH] where a set of distinct
topological charges are allowed, bearing a fractional
spin formula bilinear in the charges.
Interestingly, 
such a formula, which is an analogue of the $Q^2$-formula
mentioned above,
seems to be the norm for soliton solutions
in various $2 + 1$ dimensional
models which exhibit fractional spin (see, for example, [\JW]).
We are thus more than curious to see whether the bilinear 
formula holds universally in 
$2 + 1$ dimensions and, if so, to find a topological cause for the 
universality.  We hope to answer this question in our future 
publications.

\bigskip

\noindent
{\bf Acknowledgements:}
We are grateful to H.~Otsu and S.~Tanimura for helpful 
discussions and K.~Horie and M.~Barton for valuable comments.
This work is supported in part by the Grant-in-Aid for Scientific
Research from the Ministry of Education, Science and Culture
(No.199707529).

\vfill\eject

  \vfill\eject\immediate\closeout\reffile
  \centerline{{\bf References}}\bigskip\frenchspacing%
  \input refs.tmp\vfill\eject\nonfrenchspacing

\bye